# Proving that $P \neq NP$ and $P \subset NP \cap co-NP$

Ron A. Cohen   Oct. 19, 2004

E-mail: cohen07@bezeqint.net

## ABSTRACT

*The open question, $P \stackrel{?}{=} NP$, was presented by Cook (1971) [1]. In this paper, a proof that $P \neq NP$ is presented. In addition, it is shown that $P \subset NP \cap co-NP$. Finally, the exact inclusion relationships between the classes P, NP and co-NP are presented (see Fig. 3).*

## NOTATION

| | |
|---|---|
| D | - Deterministic Turing machine [2]. |
| ND | - Nondeterministic Turing machine [2]. |
| $D_{new}$ | - A new deterministic machine that is introduced in this paper. |
| $ND_{new}$ | - A new nondeterministic machine that is introduced as well. |
| $\equiv$ | - Polynomial equivalence between two computing machines, in the sense that each problem that can be solved in polynomial time by one machine, can be solved in polynomial time by the other. |
| $\sum = \{0,1\}$ | - Our alphabet. |

## INTRODUCTION

In order to prove that $P \neq NP$ (see Fig. 1), it is sufficient to prove that the machine D is not polynomially equivalent to the machine ND. First, a new deterministic machine, $D_{new}$, is presented, and so is its nondeterministic version, $ND_{new}$. Then it is proven that D is polynomially equivalent to $D_{new}$ (1) and that ND is polynomially equivalent to $ND_{new}$ (2). Afterwards, it is proven that $D_{new}$ is not polynomially equivalent to $ND_{new}$ (3). From (1), (2) and (3), it is concluded that D is not polynomially equivalent to ND (4).

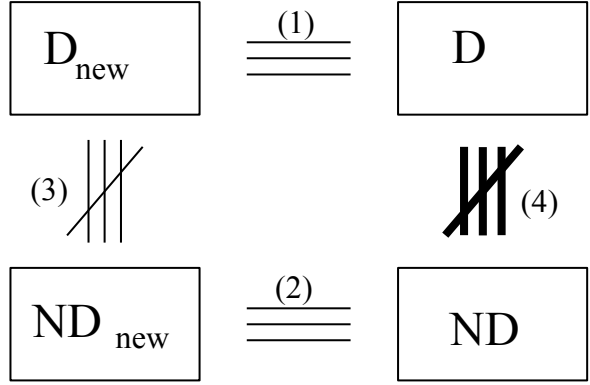

Fig 1. Scheme of $P \neq NP$ proof

Further on, it is proven that $P \subset NP \cap co-NP$. A decision problem, "A", is presented to the new machines, and the following is proven:

- "A" cannot be decided in polynomial time with $D_{new}$.
- If the answer is "yes", then there exists a polynomial time certificate.
- If the answer is "no", then a polynomial time certificate exists as well.

From (1) and (2) it follows that there is a decision problem, $\tilde{A}$, compatible to "A", which is presented to the Turing machines, such that:

- $\tilde{A}$ cannot be decided in polynomial time with D.
- If the answer is "yes", then there exists a polynomial time certificate.
- If the answer is "no", then a polynomial time certificate exists as well.

Hence- $\tilde{A} \in \bar{P} \cap NP \cap co-NP$. Therefore- $P \subset NP \cap co-NP$.

## DISSCUSION

### Description of the machine $D_{new}$

As shown in Figure 2, the machine is built from two tapes; each one is infinite in one direction only. The first tape has a read-write cursor. The second tape has a **write-only** cursor. The first tape will be called "Regular Tape", and the second one will be called, "Hidden Tape". There is a single cell with a read-write cursor that can receive either "0" or "1" as an input. If "0" is received, then the input to the problem appears on the "Regular Tape", and if "1" is received, then the input to the problem appears on the "Hidden Tape". This cell will be called "Input Button". In addition, there are two single cells. Each cell has a read-write cursor, and both cells are initialized to zero. One cell will be called "Interrupt Button", and the other- "Answer Button". "1" on the "Interrupt Button" indicates that an interrupt is required or that an interrupt is in process. "0" on the "Interrupt Button" indicates that no interrupt is required. If an interrupt is required, it is handled right away. A mechanism of the machine compares the word on the "Regular Tape", u, and the word on the "Hidden Tape", w. If $u = w$ then the machine writes "1" to the "Answer Button", and if $u \neq w$ then the machine writes "0" to the "Answer Button". In the end of an interrupt, "0" will automatically be written to the "Interrupt Button".

The input cannot be read directly from the "Hidden Tape", since its cursor is write-only. In order to reveal the word, the following steps must be made:

(1) Write a word, $u \in \Sigma^*$, on the "Regular Tape".

(2) Ask for an interrupt by writing "1" to the "Interrupt Button".

(3) If "1" appears on the "Answer Button", then the word has been revealed. Otherwise, return to (1).

Notice that the time required for an interrupt is polynomial (and actually linear) in the length of the input (|w|).

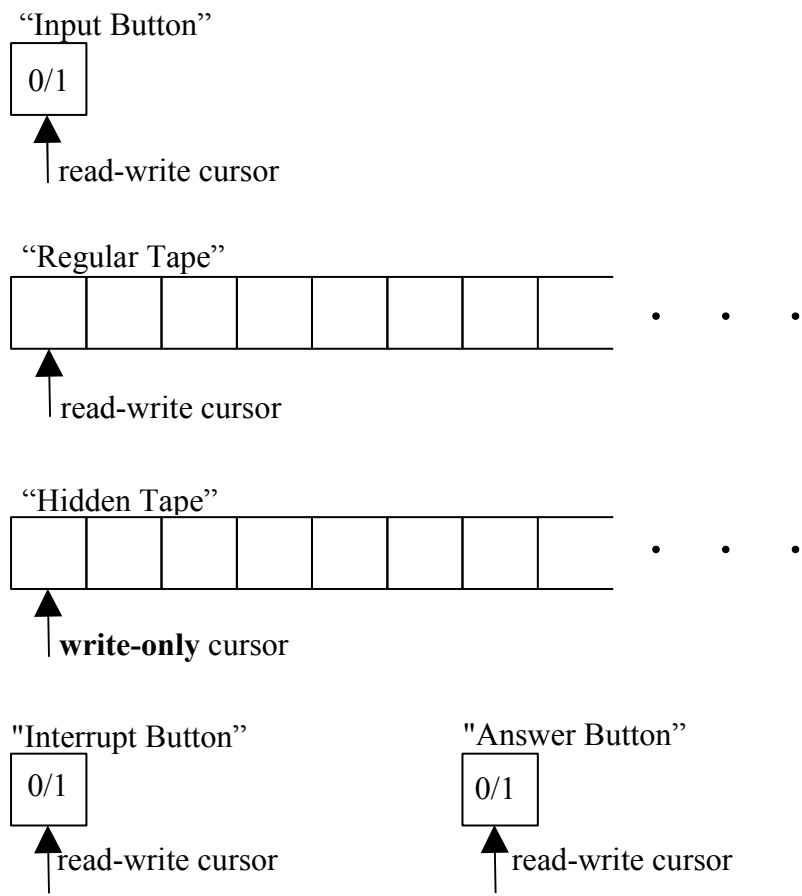

Fig 2. Scheme of the machine $D_{new}$

## Description of the machine $ND_{new}$

The difference between $D_{new}$ and $ND_{new}$ is analogous to the difference between D and ND. $ND_{new}$ is similar to $D_{new}$, except for the following difference:

> In $D_{new}$, from any given state, there could be one transition at the most. In $ND_{new}$, there could be three transitions at the most, from any given state. When trying to solve a problem with $ND_{new}$, it is assumed that the best transition is always chosen.

## Proving that P ≠ NP

Proofs of (1)-(4) from Figure 1:

(1) $D \equiv D_{new}$

We will prove:

> (a) Any problem that can be solved in polynomial time with D, can also be solved in polynomial time with $D_{new}$.
>
> (b) Any problem that can be solved in polynomial time with $D_{new}$, can also be solved in polynomial time with D.

Proofs:

> (a) Lets assume that "A" is a problem that can be solved in polynomial time with D. The problem can be solved in polynomial time with $D_{new}$ as follows:
>> The input will be presented to the "Regular Tape". The "Hidden Tape" and the interrupt mechanism will be ignored, and the solution is the same solution.
>
> (b) Let "A" be a problem that can be solved in polynomial time with $D_{new}$. Lets assume that the input is presented to the "Hidden Tape". The input cannot be read directly from the "Hidden Tape", since its cursor is write-only. In order to reveal the word, the following steps must be made:
>> (1) Write a word, $u \in \Sigma^*$, on the "Regular Tape".
>>
>> (2) Ask for an interrupt by writing "1" to the "Interrupt Button".
>>
>> (3) If "1" appears on the "Answer Button", then the word has been revealed. Otherwise, return to (1).
>
> So in the worst case, exponential time is needed, in order to decipher the "Hidden" word. Therefore, "A" is presented to the "Regular Tape", and it can be solved in polynomial time with D, in the same way.

We will comment that there are problems that their input is presented to the "Hidden Tape", yet they can be solved in polynomial time. These problems ignore the input that appears on the "Hidden Tape", and can be solved in constant time by both D and $D_{new}$.

## (2) ND ≡ $ND_{new}$

We will prove:

(a) Any problem that can be solved in polynomial time with ND can also be solved in polynomial time with $ND_{new}$.

(b) Any problem that can be solved in polynomial time with $ND_{new}$ can also be solved in polynomial time with ND.

Proofs:

(a) Lets assume that "A" is a problem that can be solved in polynomial time with ND. The problem can be solved in polynomial time with $ND_{new}$ in the following way:

> The input will be presented to the "Regular Tape". The "Hidden Tape" and the interrupt mechanism will be ignored, and the solution is the same solution.

(b) Let "A" be a problem that can be solved in polynomial time by $ND_{new}$. If "0" appears on the "Input Button", meaning that the input is presented to the "Regular Tape", then "A" can be solved in polynomial time with ND in the same manner. If "1" appears on the "Input Button", meaning that the input is presented to the "Hidden Tape", then there exists a compatible problem, $\widetilde{A}$, that is presented to ND. Lets assume, for the sake of contradiction, that there is no polynomial solution for $\widetilde{A}$ with ND. Now, if we write "0" on the "Input Button" and present the problem $\widetilde{A}$ to $ND_{new}$, then $\widetilde{A}$ does not have a polynomial time solution with $ND_{new}$. On the one hand, "A" can be solved in polynomial time with $ND_{new}$. On the other hand, $\widetilde{A}$, which is a different presentation of the problem "A", cannot be solved in polynomial time with $ND_{new}$. That is an absurd. Hence, it is concluded that "A" can always be solved in polynomial time with ND.

**(3) $D_{new} \neq ND_{new}$**

Lets define a decision problem "Q":

Input:
- "1" appears on the "Input Button".
- Nothing appears on the "Regular Tape".
- A word $w \in \Sigma^*$ appears on the "Hidden Tape".
- "0" appears on the "Interrupt Button".
- "0" appears on the "Answer Button".

Output:

"yes" $\Leftrightarrow$ There exists a word $u \in \Sigma^*$ such that $w = 1 \cdot u$     ("·" - concatenation)

In order to prove that $D_{new} \not\equiv ND_{new}$, it is sufficient to show that:
  (a) "Q" can be decided in polynomial time with $ND_{new}$.
  (b) "Q" cannot be decided in polynomial time with $D_{new}$.

Proofs:
  (a) Lets present a nondeterministic algorithm that decides "Q" in polynomial time:
    (1) Write "1" on the "Regular Tape", and move to the right.
    (2) Choose nondeterministicly between three actions:
      - Writing "0" on the "Regular Tape" and moving to the right.
      - Writing "1" on the "Regular Tape" and moving to the right.
      - Asking for an interrupt by writing "1" to the "Interrupt Button". After the interrupt request, if the answer is positive then halt at the "yes" state; otherwise halt at the "no" state.
  
  The time needed is polynomial in the length of the input (|w|), and actually linear.

(b) Any algorithm that tries to solve the problem, must try to reveal the first letter on the "Hidden Tape". In order to reveal a single letter, the whole word must be revealed. We do not have any preliminary knowledge concerning the word that appears on the "Hidden Tape". Therefore the following steps must be made:

(1) Write a word, $u \in \Sigma^*$, on the "Regular Tape".

(2) Ask for an interrupt by writing "1" to the "Interrupt Button".

(3) If "1" appears on the "Answer Button", then the word has been revealed. Otherwise, return to (1).

That is why all algorithms must check $\Omega(2^{|w|})$ words at the worst case. Hence, there is no algorithm that decides Q in polynomial time.

(4) D $\not\equiv$ ND, or in other words- $P \neq NP$

From (1), (2) and (3) it follows that (4) is true, since if D was polynomially equivalent to ND, then we would conclude that D is not polynomially equivalent to itself, and that is an absurd.

**Proving that P $\subset$ NP $\cap$ co-NP**

Lets reconsider the problem "Q":

Input:
- "1" appears on the "Input Button".
- Nothing appears on the "Regular Tape".
- A word $w \in \Sigma^*$ appears on the "Hidden Tape".
- "0" appears on the "Interrupt Button".
- "0" appears on the "Answer Button".

Output:

"yes" $\Leftrightarrow$ There exists a word $u \in \Sigma^*$ such that $w = 1 \cdot u$    ("·" - concatenation)

We will prove:

(a) "Q" cannot be decided in polynomial time with $D_{new}$.

(b) If the answer is "yes", then a polynomial certificate exists.

(c) If the answer is "no", then a polynomial certificate exists.

Proofs:

(a) Any algorithm that tries to solve the problem, must try to reveal the first letter on the "Hidden Tape". In order to reveal a single letter, the whole word must be revealed. We do not have any preliminary knowledge concerning the word that appears on the "Hidden Tape". Therefore, there is no alternative, but to choose words from $\Sigma^*$, write them on the "Regular Tape" and ask for an interrupt. That is why all algorithms must check at least $\Omega(2^{|w|})$ words at the worst case. Hence, there is no algorithm that decides "Q" in polynomial time.

(b) The certificate:
- Write the word w on the "Regular Tape".
- Ask for an interrupt and receive a positive answer.
- Make sure that w starts with "1".

The above certification requires polynomial time, and actually linear time.

(c) The certificate:
- Write the word w on the "Regular Tape".
- Ask for an interrupt and receive a positive answer.
- Make sure that w starts with "0".

The above certification requires linear time.

From (1) and (2) (from Fig. 1) and from (a)-(c), it follows that there is a problem in $\overline{P} \cap NP \cap co-NP$. Therefore $P \subset NP \cap co-NP$.

## CONCLUSIONS

In this paper, a proof that $P \neq NP$ is presented. Further more, it was shown that $P \subset NP \cap co-NP$.

It is possible to prove, in a similar way, that there exist problems R and S such that

$$R \in NP \cap \overline{co-NP} \cap \overline{P}$$

And $\quad S \in co-NP \cap \overline{NP} \cap \overline{P}$

The exact inclusion relationships between the classes P, NP and co-NP are shown in Figure 3.

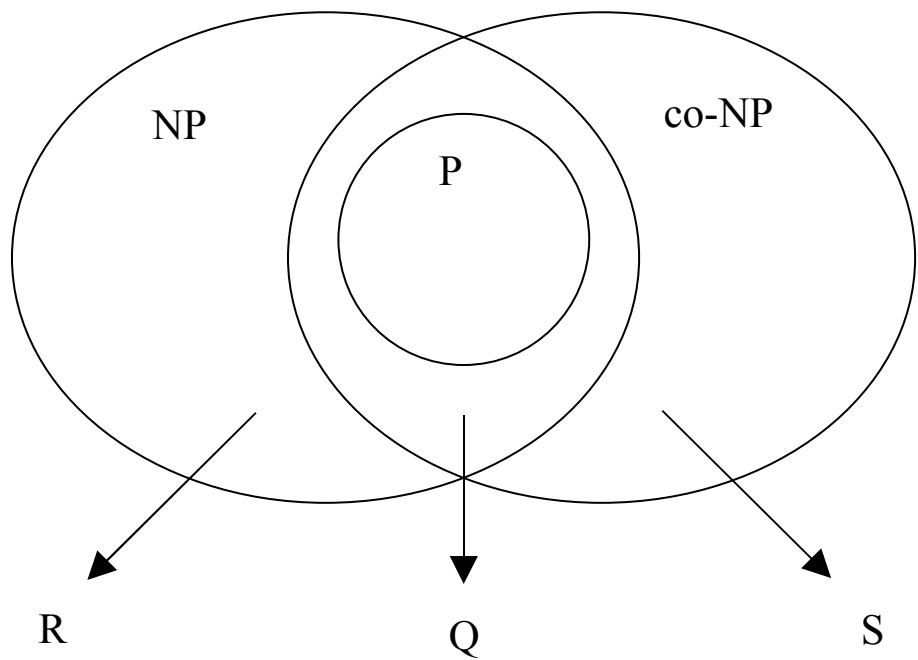

Fig 3. The exact inclusion relationships between the classes P, NP and co-NP

## ACKNOWLEDGMENTS

The author is grateful to his parents.